# Polarized Nucleon Structure Functions from Lattice QCD

M. Göckeler[1,2] [*], R. Horsley[3], E.-M. Ilgenfritz[3], H. Perlt[4],
P. Rakow[5], G. Schierholz[6,1] and A. Schiller[4]

[1] HLRZ, c/o Forschungszentrum Jülich,
D-52425 Jülich, Germany

[2] Institut für Theoretische Physik, RWTH Aachen
D-52056 Aachen, Germany

[3] Institut für Physik, Humboldt- Universität,
D-10115 Berlin, Germany

[4] Fakultät für Physik und Geowissenschaften,
Universität Leipzig,
D-04109 Leipzig, Germany

[5] Institut für Theoretische Physik, Freie Universität,
D-14195 Berlin, Germany

[6] Deutsches Elektronen-Synchrotron DESY,
D-22603 Hamburg, Germany

## Abstract

We describe a high statistics quenched lattice QCD calculation of the moments of the polarized deep-inelastic structure functions $g_1$ and $g_2$ of the proton and neutron.

## 1 Introduction

The polarized nucleon structure functions $g_1$ and $g_2$ have attracted a lot of interest in the last years. Especially the measurements of $g_1$ [1] provoked a lively discussion under the headline "proton spin crisis". The structure function $g_2$, on the other hand, offers the first possibility of a direct measurement of a higher twist (=3) operator matrix element [2]. Experiments that perform measurements of $g_2$ are being undertaken at DESY (HERA) and SLAC. On the theoretical side it is widely believed that quantum chromodynamics (QCD) is the correct theory of the strong interactions. To put this belief on a solid experimental footing one has to compare the experimental results with the predictions from QCD. It is therefore a major challenge for theoretical particle physics to find answers to the question "What does QCD tell us about the structure functions in deep-inelastic lepton nucleon scattering?"

---

[*] Speaker at the workshop

Asymptotic freedom allows us to calculate the violations of Bjorken scaling in the (moments of the) structure functions using perturbation theory and Wilson's operator product expansion (OPE). However, the computation of the structure functions themselves (for a fixed value of the momentum transfer $Q^2$) requires a nonperturbative method. This talk describes an attempt [3] to perform such a calculation applying lattice gauge theory and Monte Carlo simulations. (For earlier work see ref. [4].) We have studied polarized as well as unpolarized structure functions, but in accordance with the subject of the workshop we shall restrict ourselves to the polarized case in this talk.

From the OPE we have for the leading twist contributions in the chiral limit:

$$
\begin{aligned}
2\int_0^1 dx\, x^n g_1(x, Q^2) &= \frac{1}{2} \sum_{f=u,d} e_{1,n}^{(f)}(\mu^2/Q^2, g(\mu)) a_n^{(f)}(\mu), \\
2\int_0^1 dx\, x^n g_2(x, Q^2) &= \frac{1}{2} \frac{n}{n+1} \sum_{f=u,d} \left[ e_{2,n}^{(f)}(\mu^2/Q^2, g(\mu)) d_n^{(f)}(\mu) \right. \\
&\qquad\qquad\qquad\qquad \left. - e_{1,n}^{(f)}(\mu^2/Q^2, g(\mu)) a_n^{(f)}(\mu) \right]
\end{aligned}
\qquad (1)
$$

for even $n$ and $n \geq 0$ ($n \geq 2$) for $g_1$ ($g_2$). The reduced matrix elements $a_n^{(f)}$, $d_n^{(f)}$ for the flavors $f = u, d$ are defined in terms of forward nucleon matrix elements of the operators

$$
\mathcal{O}_{\sigma\mu_1\cdots\mu_n}^{5(f)} = \left(\frac{i}{2}\right)^n \bar{\psi} \gamma_\sigma \gamma_5 \overleftrightarrow{D}_{\mu_1} \cdots \overleftrightarrow{D}_{\mu_n} \psi - \text{traces} \qquad (2)
$$

with $\psi = u(d)$ for $f = u(d)$ according to (see e.g. ref. [2])

$$
\langle p, s | \mathcal{O}_{\{\sigma\mu_1\cdots\mu_n\}}^{5(f)} | p, s \rangle = \frac{a_n^{(f)}}{n+1} [s_\sigma p_{\mu_1} \cdots p_{\mu_n} + \cdots - \text{traces}], \qquad (3)
$$

$$
\langle p, s | \mathcal{O}_{[\sigma\{\mu_1]\cdots\mu_n\}}^{5(f)} | p, s \rangle = \frac{d_n^{(f)}}{n+1} [(s_\sigma p_{\mu_1} - s_{\mu_1} p_\sigma) p_{\mu_2} \cdots p_{\mu_n} + \cdots - \text{traces}]. \qquad (4)
$$

Here $\{\cdots\}$ and $[\cdots]$ denote symmetrization and antisymmetrization, respectively, and $\mu$ is the renormalization scale at which the operators are defined. In the flavor singlet case purely gluonic operators will contribute also.

The Wilson coefficients $e_{j,n}^{(f)}(\mu^2/Q^2, g(\mu))$, $j = 1, 2$, are calculated in perturbation theory. For electroproduction they have the form

$$e_{j,n}^{(f)}(\mu^2/Q^2, g(\mu)) = Q^{(f)2}(1 + g(\mu)^2 \bar{e}_{j,n}(\mu^2/Q^2, g(\mu))), \tag{5}$$

where $Q^{(f)}$ are the quark charges. The nucleon states $|p, s\rangle$ with momentum $p$ and covariant spin vector $s$ are normalized according to

$$\langle p, s | p', s' \rangle = (2\pi)^3 \cdot 2E_p \delta(\vec{p} - \vec{p}') \delta_{s,s'} \tag{6}$$

where $s^2 = -m_N^2$ ($m_N$ = nucleon mass). Note that $\mathcal{O}_{\{\sigma\mu_1\cdots\mu_n\}}^{5(f)}$ has twist 2, whereas $\mathcal{O}_{[\sigma\{\mu_1]\cdots\mu_n\}}^{5(f)}$ is an operator of twist 3. Hence, in the moments of $g_2$, $a_n^{(f)}$ represents the so-called Wandzura-Wilczek contribution [5].

The basic task is now the computation of the matrix elements (3), (4) in a polarized nucleon. Depending on the choice of the indices $\sigma, \mu_1, \ldots, \mu_n$ it will be necessary to give the nucleon a nonvanishing three-momentum.

## 2 Lattice Calculation

How are these matrix elements calculated by lattice Monte Carlo simulations?

Analytic continuation to imaginary time, i.e. putting $x_0 = -\mathrm{i} x_4$, leads us from Minkowski space to Euclidean space and allows us to interpret the functional integral of a quantum field theory as the ensemble average of a statistical mechanical system in four dimensions. We can then take advantage of the experience gained in statistical mechanics, in particular we may discretize the Euclidean space and use Monte Carlo simulations for the calculation of the expectation values we are interested in.

Roughly speaking, the continuation to imaginary time means that the unitary time evolution operator $\mathrm{e}^{-\mathrm{i}Ht}$ is replaced by the self-adjoint operator $\mathrm{e}^{-Ht}$. Hence the exponential decay of Euclidean correlation functions, the so-called Schwinger functions, contains information about the energy spectrum of the theory, and the amplitudes multiplying the exponentials are related to matrix elements.

To be more specific, consider the two-point function

$$\langle 0 | B(t_2) \bar{B}(t_1) | 0 \rangle = \langle 0 | \mathrm{e}^{\mathrm{i}Ht_2} B(0) \mathrm{e}^{-\mathrm{i}Ht_2} \mathrm{e}^{\mathrm{i}Ht_1} \bar{B}(0) \mathrm{e}^{-\mathrm{i}Ht_1} | 0 \rangle$$
$$= \langle 0 | B(0) \mathrm{e}^{-\mathrm{i}H(t_2-t_1)} \bar{B}(0) | 0 \rangle \tag{7}$$

with suitable interpolating fields $B$, $\bar{B}$ for the nucleon. Analytic continuation yields the correlation function

$$\langle 0 | B \mathrm{e}^{-Ht} \bar{B} | 0 \rangle = \langle B(t) \bar{B}(0) \rangle. \tag{8}$$

As $t$ becomes large, it is dominated by the lowest-lying state coupling to $B$, i.e. the nucleon with energy $E_N$:

$$\langle B(t)\bar{B}(0)\rangle = \langle 0|B|N\rangle \mathrm{e}^{-E_N t}\langle N|\bar{B}|0\rangle + \cdots \tag{9}$$

In order to suppress the contributions of higher excitations indicated by the dots in (9) we use the freedom in the choice of the interpolating field $B$ by applying, e.g., 'smearing' techniques [3].

For three-point functions we have similarly

$$\langle 0|B(t)\mathcal{O}(\tau)\bar{B}(0)|0\rangle = \langle 0|B(0)\mathrm{e}^{-\mathrm{i}H(t-\tau)}\mathcal{O}(0)\mathrm{e}^{-\mathrm{i}H\tau}\bar{B}(0)|0\rangle \tag{10}$$

In Euclidean space we obtain for $t > \tau$ the correlation function

$$\langle 0|B\mathrm{e}^{-H(t-\tau)}\mathcal{O}\mathrm{e}^{-H\tau}\bar{B}|0\rangle = \langle B(t)\mathcal{O}(\tau)\bar{B}(0)\rangle \tag{11}$$

and in the limit of large time differences $t - \tau$, $\tau$ we have

$$\begin{aligned}\langle B(t)\mathcal{O}(\tau)\bar{B}(0)\rangle &= \langle 0|B|N\rangle \mathrm{e}^{-E_N(t-\tau)}\langle N|\mathcal{O}|N\rangle \mathrm{e}^{-E_N\tau}\langle N|\bar{B}|0\rangle + \cdots \\ &= \langle 0|B|N\rangle \mathrm{e}^{-E_N t}\langle N|\bar{B}|0\rangle\langle N|\mathcal{O}|N\rangle + \cdots\end{aligned} \tag{12}$$

Hence the leading contribution is independent of $\tau$ and we can determine the desired nucleon matrix elements from ratios of three-point functions to two-point functions:

$$\frac{\langle B(t)\mathcal{O}(\tau)\bar{B}(0)\rangle}{\langle B(t)\bar{B}(0)\rangle} = \langle N|\mathcal{O}|N\rangle + \cdots \text{ for } 0 \ll \tau \ll t\,. \tag{13}$$

In order to regularize the ultraviolet divergences we introduce a regular hypercubic lattice with lattice spacing $a$. Our lattice has the size $16^3 \times 32$, with the larger extent corresponding to (Euclidean) time. The quark fields are chosen periodic in space and antiperiodic in time, whereas the gauge fields are periodic in all directions. We take the standard Wilson action for the gauge fields and use Wilson fermions to put the quarks on the lattice. The latter choice implies an explicit breaking of chiral symmetry by lattice artifacts and the bare quark mass is replaced by the so-called hopping parameter $\kappa$. The quark mass is then approximately given by

$$m_q a \approx 0.56 \left(\frac{1}{\kappa} - \frac{1}{\kappa_c}\right)\,, \tag{14}$$

where the critical hopping parameter $\kappa_c$ corresponds to the chiral limit (vanishing pion mass). Our bare gauge coupling is $g = 1$ and we have data for three values of the hopping parameter, $\kappa = 0.155$, $0.153$ and $0.1515$, corresponding to quark masses $m_q$ of roughly 70, 130 and 190 MeV, respectively. Hence we can try to extrapolate our results to physical quark masses. For the accuracy we are aiming at this is the same as extrapolating to the chiral limit.

Our calculations are performed in the so-called "quenched" or "valence" approximation, i.e. we neglect all internal quark loops, as this drastically reduces the necessary computer time. Correspondingly, we disregard in the operator insertion the so-called "disconnected" contribution where the operator is not directly coupled to the valence quarks in the nucleon. We are, however, trying to calculate this contribution thus making a first step towards incorporating sea quark effects, but we have no results yet.

## 3 Operators and Matrix Elements

In Euclidean space we deal with operators of the form

$$\mathcal{O}^5_{\sigma\mu_1\cdots\mu_n} = \bar{\psi}\gamma_\sigma\gamma_5 D_{\mu_1}\cdots D_{\mu_n}\psi - \text{traces}\,. \tag{15}$$

On the lattice, $D_\mu$ is of course the lattice covariant derivative. More explicitly, we study the three operators

$$\mathcal{O}^5_2 = \bar{\psi}\gamma_2\gamma_5\psi \text{ for } a_0\,, \tag{16}$$

$$\mathcal{O}^5_{\{214\}} = \frac{1}{3!}\left(\mathcal{O}^5_{214} + \mathcal{O}^5_{241} + \mathcal{O}^5_{124} + \mathcal{O}^5_{142} + \mathcal{O}^5_{412} + \mathcal{O}^5_{421}\right) \text{ for } a_2\,, \tag{17}$$

$$\mathcal{O}^5_{[2\{1]4\}} = \frac{1}{3}\left(\mathcal{O}^5_{214} + \mathcal{O}^5_{241} - \tfrac{1}{2}\mathcal{O}^5_{124} - \tfrac{1}{2}\mathcal{O}^5_{142} - \tfrac{1}{2}\mathcal{O}^5_{412} - \tfrac{1}{2}\mathcal{O}^5_{421}\right) \text{ for } d_2\,. \tag{18}$$

Two criteria guided us towards the above choice of operators. First, we do not want to be forced to have more than one component of $\vec{p}$ nonzero. For this would lead to higher energies of the nucleon states to be considered, hence to a faster decrease of the propagator and ultimately to a worse signal-to-noise ratio.

Secondly, we have to avoid mixing with lower-dimensional operators. These would come with coefficients diverging like a power of $a^{-1}$. So the necessary subtractions would be large and furthermore uncalculable in perturbation theory [6]. Note that the mixing pattern is much more complicated on the lattice than it is in the continuum, because the constraints imposed by symmetry are less stringent. Instead of covariance with respect to the orthogonal group O(4) we only have at our disposal the finite subgroup H(4) (the so-called hypercubic group) which leaves the lattice invariant. Multiplets that are irreducible with respect to O(4) may become reducible under H(4), and the possibilities for mixing increase.

Corresponding to the above operators we choose the polarization in the 2-direction and consider the two momenta $\vec{p} = \vec{0}$, $(2\pi/16, 0, 0)\cdot 1/a$. Furthermore, we have to choose a fixed value for $t$ in eq.(13). This should be as large as possible in order to allow for a long "plateau", i.e. an interval in $\tau$ where the ratio (13) is independent of $\tau$ and hence enables a reliable determination of the desired matrix element. We have taken $t = 13$. The ratios pertaining to the reduced matrix elements $a_0$, $a_2$, and $d_2$ are shown in Fig.1.

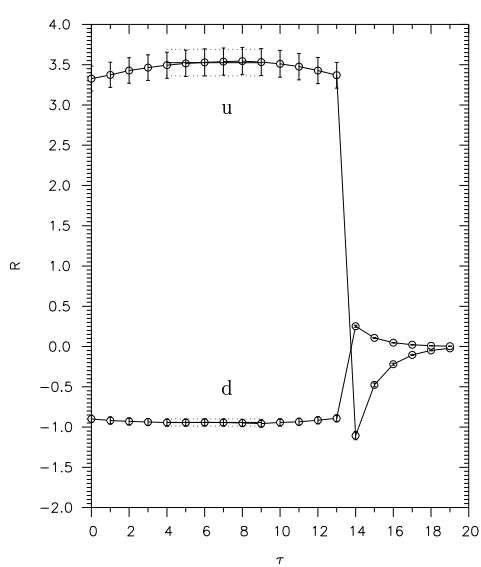
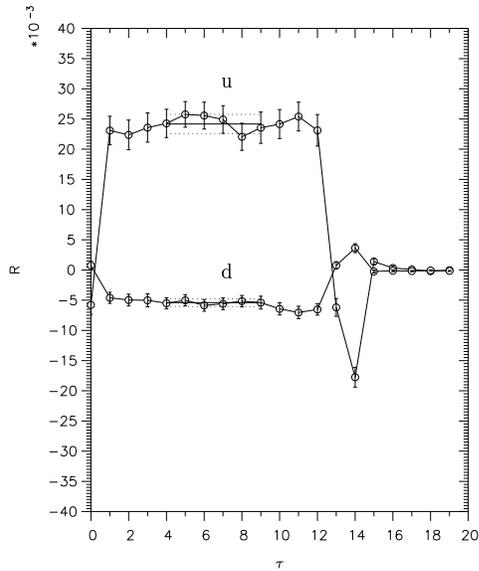
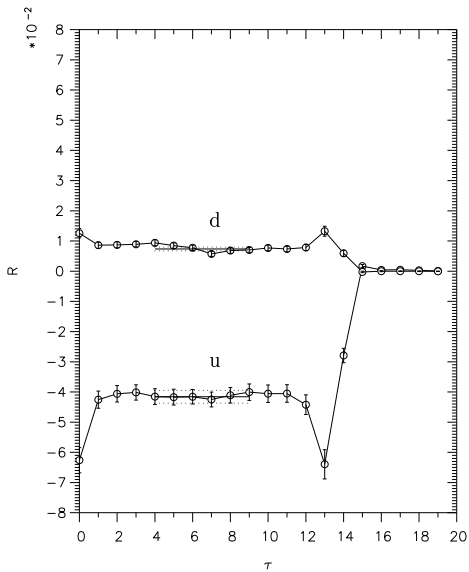

**Figure 1:** The ratio $R$ for $u$ and $d$ quark insertions for $\kappa = 0.153$. (a) $-iR_{a_0}$, (b) $R_{a_2}$ and (c) $R_{d_2}$. $\tau$ is given in lattice units. The horizontal lines indicate the fitted value of $R$ on the plateau together with its error and the fit range.

For the three operators (16)–(18) the ratios $R$ of three-point functions over two-point functions (in lattice units) can be written as

$$R_{a_0} = \frac{i}{Z_{a_0}} \frac{1}{2\kappa} \frac{1}{2} a_0 \text{ for } \vec{p} = \vec{0}, \qquad (19)$$

$$R_{a_2} = \frac{1}{Z_{a_2}} \frac{1}{2\kappa} \frac{m_N p_1}{6} a_2, \qquad (20)$$

$$R_{d_2} = \frac{1}{Z_{d_2}} \frac{1}{2\kappa} \frac{m_N p_1}{3} d_2. \qquad (21)$$

The factor $2\kappa$ results from the different normalization of the lattice quark fields as compared to the continuum conventions. The $Z$'s ($Z = Z(a\mu)$) are the renormalization factors needed to construct operators renormalized at a scale $\mu$ from the bare lattice operators. At present, we take the $Z$'s defined according to the momentum subtraction scheme from a one-loop perturbative calculation in the chiral limit. (A nonperturbative calculation seems however possible [7] and is being done.) If necessary, we can convert the numbers to the $\overline{\text{MS}}$ scheme, in which the Wilson coefficients are usually calculated. In the following we shall quote our results extrapolated to the chiral limit for

$$\mu^2 = Q^2 = a^{-2} \approx 5 \text{GeV}^2 \,, \tag{22}$$

which eliminates the logarithms in the Wilson coefficients and renormalization constants.

Let us note in passing that the operator (18) used for the determination of $d_2$ could mix on the lattice with some other operator of the same dimension. However, the matrix element of the latter operator turns out to be small and the mixing coefficient vanishes in one-loop perturbation theory. Hence we have neglected the mixing.

## 4 Discussion of Results

In Fig.2 we show our results and their extrapolation to the chiral limit. Instead of $a_0^{(f)}$ we plot $\Delta u$ and $\Delta d$ defined by

$$a_0^{(u)} = 2\Delta u \,, \ a_0^{(d)} = 2\Delta d \tag{23}$$

and interpreted as the fraction of the nucleon spin that is carried by the $u$ and $d$ quarks, respectively. For the proton we find in the mom scheme

$$\Delta u = 0.83(7)\,, \ \Delta d = -0.24(2)\,. \tag{24}$$

Since these numbers were obtained in the quenched approximation, they should be considered as an estimate of the valence quark contribution only. A quantity which is expected to be less sensitive to sea quark effects is the difference $\Delta u - \Delta d$ for which we get in the proton:

$$\Delta u - \Delta d = 1.07(9)\,. \tag{25}$$

This is to be compared with the experimental value of the axial vector coupling constant of the nucleon, which is 1.26.

Instead of looking for observables which are dominated by the valence quarks one can try to incorporate at least part of the sea quark effects by taking into account the "disconnected" contribution to the matrix element. Our calculations

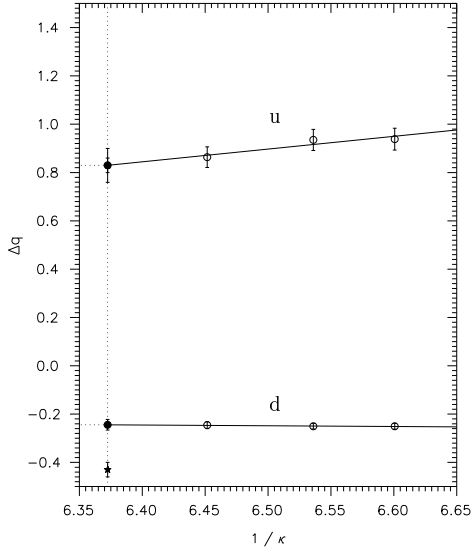
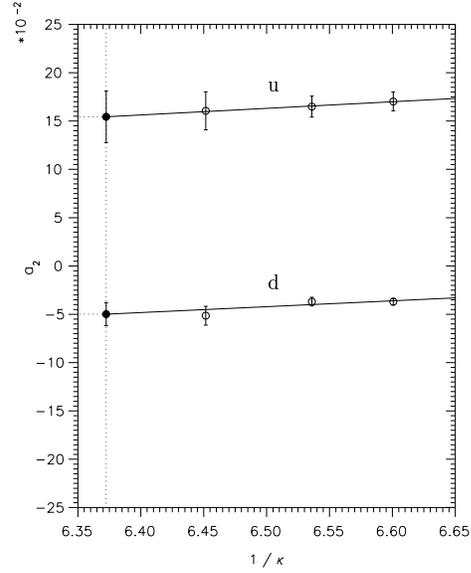
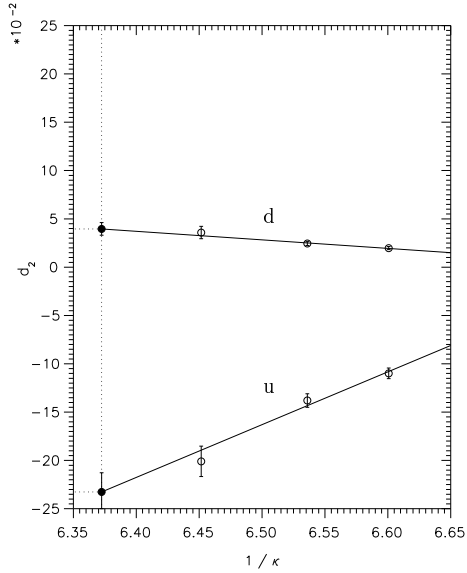

**Figure 2:** $\Delta q$, $a_2$, and $d_2$ for the proton as a function of $1/\kappa$, together with a linear fit to the data. The solid symbols indicate the extrapolation to the chiral limit. In (a) we compare our numbers with the phenomenological values of Ref.[9], which are marked by asterisks. (The value for $\Delta u$ is hidden behind the lattice number.)

of this quantity are not yet finished, but for $\Delta q$ we can use the results of two other groups [8]. Renormalizing these results according to our conventions we obtain in total

|  | $\Delta u$ | $\Delta d$ | $\Delta s$ | $\Delta\Sigma$ |
|---|---|---|---|---|
| Lattice QCD | 0.69(9) | −0.38(5) | −0.13(4) | 0.18(11) |
| EK | 0.83(3) | −0.43(3) | −0.10(3) | 0.31(07) |

For comparison we show in the row labeled EK the results of the recent phenomenological analysis by Ellis and Karliner [9] using $Q^2 = 10\,\text{GeV}^2$.

Rewriting our results in terms of the lowest moment of $g_1$ we get (cf. eq.(1))

$$\int_0^1 dx\, g_1(x, Q^2) = \begin{cases} 0.166(16) & \text{proton,} \\ -0.008(09) & \text{neutron,} \end{cases} \quad (26)$$

with the "disconnected" contributions left out. This omission should be less harmful in the difference of proton and neutron structure functions. Indeed, our result

$$\int_0^1 dx(g_1^p(x, Q^2) - g_1^n(x, Q^2)) = 0.174(25) \quad (27)$$

compares rather favorably with the experimental value of $0.161\pm 0.007\pm 0.015$ [9].

In the higher moments of $g_1$ sea quark effects are expected to be suppressed. From our calculation of $a_2$ we obtain in the chiral limit

$$\int_0^1 dx\, x^2 g_1(x, Q^2) = \begin{cases} 0.0150(32) & \text{proton,} \\ -0.0012(20) & \text{neutron,} \end{cases} \quad (28)$$

consistent with experiment.

Let us finally come to the structure function $g_2$. Extrapolating our results for $a_2$ and $d_2$ to the chiral limit we get

$$\int_0^1 dx\, x^2 g_2(x, Q^2) = \begin{cases} -0.0161(16) & -0.0100(22) = -0.0261(38) & \text{proton,} \\ -0.0013(09) & +0.0009(13) = -0.0004(22) & \text{neutron.} \end{cases} \quad (29)$$

Here the first number comes from $d_2$, while the second number comes from $a_2$. Hence the twist-three operator provides a significant contribution, at least in the case of the proton. This result disagrees with sum rule calculations [10], which suggest that for the proton $d_2$ is very small.

## 5 Conclusion

We have presented a calculation of moments of the polarized nucleon structure functions from first principles. The results show that such a calculation is feasible, at least for the lower moments. We had to make some approximations though, e.g. we have neglected sea quarks and gluonic operators. Furthermore, there are systematic uncertainties due to the extrapolation to physical quark masses, cutoff effects etc. However, all these shortcomings can (and will) be improved. But even at the present stage the agreement with the experimental data, as far as they exist, is rather satisfactory. More precise experiments measuring $g_2$ should reveal if the twist-three contribution indeed plays such an important role as our results suggest.

## Acknowledgments

This work was supported in part by the Deutsche Forschungsgemeinschaft. The numerical calculations were performed on the Quadrics parallel computers at Bielefeld University and at DESY (Zeuthen). We wish to thank both institutions for their support and in particular the system managers M. Plagge and H. Simma for their help.

## References

[1] J. Ashman et al., Phys. Lett. **B206** (1988) 364; Nucl. Phys. **B238** (1990) 1; D. L. Anthony et al., Phys. Rev. Lett. **71** (1993) 959; B. Adeva et al., Phys. Lett. **B302** (1993) 533; D. Adams et al., Phys. Lett. **B329** (1994) 399; Phys. Lett. **B339** (1994) 332 (E); K. Abe et al., Phys. Rev. Lett. **74** (1995) 346; Phys. Rev. Lett. **75** (1995) 25.

[2] R. L. Jaffe, Comm. Nucl. Part. Phys. **19** (1990) 239.

[3] M. Göckeler, R. Horsley, E.-M. Ilgenfritz, H. Perlt, P. Rakow, G. Schierholz and A. Schiller, Nucl. Phys. **B** (Proc. Suppl.) **42** (1995) 337; Preprint DESY 95-128, HLRZ 95-36, HUB-EP-95/9 (hep-lat/9508004).

[4] G. Martinelli and C. T. Sachrajda, Nucl. Phys. **B316** (1989) 355; G. Martinelli, Nucl. Phys. **B** (Proc. Suppl.) **9** (1989) 134.

[5] S. Wandzura and F. Wilczek, Phys. Lett. **B72** (1977) 195.

[6] C. T. Sachrajda, Nucl. Phys. **B** (Proc. Suppl.) **9** (1989) 121.

[7] G. Martinelli, C. Pittori, C. T. Sachrajda, M. Testa and A. Vladikas, Nucl. Phys. **B445** (1995) 81.

[8] M. Fukugita, Y. Kuramashi, M. Okawa and A. Ukawa, Phys. Rev. Lett. **75** (1995) 2092; S. J. Dong, J.-F. Lagaë and K.-F. Liu, Phys. Rev. Lett. **75** (1995) 2096.

[9] J. Ellis and M. Karliner, Phys. Lett. **B341** (1995) 397; see also M. Karliner's contribution to this workshop.

[10] I. I. Balitsky, V. M. Braun and A. V. Kolesnichenko, Phys. Lett. **B242** (1990) 245; E. Stein, P. Górnicki, L. Mankiewicz, A. Schäfer and W. Greiner, Phys. Lett. **B343** (1995) 369.